\documentclass[conference]{IEEEtran}
\IEEEoverridecommandlockouts
\usepackage{float}  
\usepackage{cite}
\usepackage{amsmath,amssymb,amsfonts}
\usepackage{graphicx}
\usepackage{textcomp}
\usepackage{xcolor}
\usepackage{xcolor}
\definecolor{myblack}{named}{black}
\usepackage{algorithm}
\usepackage{algpseudocode}

\usepackage{hyperref,xcolor}
\definecolor{mypink}{RGB}{255, 20, 147}
\usepackage{fancyhdr}
\hypersetup{
	pdfborder={0 0 0},
    colorlinks=true,
     citecolor=black,       
     urlcolor=myblack
}

\def\BibTeX{{\rm B\kern-.05em{\sc i\kern-.025em b}\kern-.08em
    T\kern-.1667em\lower.7ex\hbox{E}\kern-.125emX}}
\begin{document}

\title{Mal-D2GAN: Double-Detector based GAN for Malware Generation\\

\thanks{
----------------------------------------------------}
}

\author{\IEEEauthorblockN{Nam Hoang Thanh}
\IEEEauthorblockA{{Academy of Cryptography Techniques} \\
Hanoi, Vietnam \\
hoangthanhnam@actvn.edu.vn}
\and
\IEEEauthorblockN{Trung Pham Duy}
\IEEEauthorblockA{{Academy of Cryptography Techniques} \\
Hanoi, Vietnam \\
trungpd@actvn.edu.vn}
\and
\IEEEauthorblockN{Lam Bui Thu\textsuperscript{*}}
\IEEEauthorblockA{{Academy of Cryptography Techniques} \\
Hanoi, Vietnam \\
lambt@actvn.edu.vn}

}

\maketitle
\begingroup
\renewcommand{\thefootnote}{\fnsymbol{footnote}}
\footnotetext{\hspace{-0.1cm}\textbf{Lam Bui Thu is the corresponding author}.} 
\endgroup

\begin{abstract}
Machine learning (ML) has been developed to detect malware in recent years. Most researchers focused their efforts on improving the detection performance but ignored the robustness of the ML models. In addition, many machine learning algorithms are very vulnerable to intentional attacks. To solve these problems, adversarial malware examples are generated by GANs to enhance the robustness of the malware detector. However, since current GAN models suffer from limitations such as unstable training and weak adversarial examples, we propose the Mal-D2GAN model to address these problems. Specifically, the Mal-D2GAN architecture was designed with double-detector and a least square loss function and tested on a dataset of 20,000 samples. The results show that the Mal-D2GAN model reduced the detection accuracy (true positive rate) in 8 malware detectors. The performance was then compared with that of the existing MalGAN and Mal- LSGAN models.
\end{abstract}


\section{Introduction}
Malware, or malicious software, refers to programs or codes intentionally designed to cause harm to personal computers, servers, or computer networks. The primary purpose of malware is to execute unauthorized actions that can lead to various illicit activities. Therefore, malware detection is an important step in the incident response process for any organization, system, or enterprise. Malware detection is divided into two main types: signature-based detection and behavior-based detection \cite{1}

Signature-based detection is a technique that compares files or network traffic with a database of known malware signatures. A signature is a unique pattern or fingerprint of bytes that identifies a specific malware. Signature-based detection can quickly and accurately detect known malware variants, but it can not detect new or unknown malware with no signature in the database. In contrast, behavior-based detection can detect new or unknown malware with no signature and malware that tries to hide or disguise its signature. Behavior-based detection is a technique that analyzes the actions or behavior of files or network traffic to detect malware. Behavior-based detection does not rely on predefined signatures, but on heuristic rules or machine- learning algorithms that can identify suspicious or malicious activities \cite{2}.

Machine learning \cite{3} can learn from the behavior of software, not just rely on pre-known signatures. This allows the detection of new variants of malware without frequent updates to signatures. However, machine learning-based malware detection systems face the drawback of susceptibility to adversarial malware samples \cite{4}. Adversarial samples generated from Generative Adversarial Networks (GANs) \cite{5} have led to numerous adversarial attacks on machine learning-based malware detection systems \cite{6}. To create adversarial malware examples to bypass ML based malware detector, Hu et al \cite{7} proposed the MalGAN model which uses adversarial samples to bypass malware detectors (a.k.a black-box detectors). The purpose of this work to enhance the robustness of malware detector. In studies such as Mal-LSGAN \cite{8} and LSGAN- AT \cite{9}, improvements using integrated activation functions were introduced to enhance the ability to bypass black-box detectors. However, recent GAN models suffer from limitations such as unstable training and weak adversarial examples. In this paper, we propose the Mal-D2GAN model with double-detector and a least square loss function and test on a dataset of 20,000 samples. The experimental results show that adversarial examples generated by our proposed model can bypass malware black-box detectors with 8 common machine learning algorithms. The proposed method’s results are higher than the MalGAN and Mal- LSGAN models proposed.

The rest of the paper is organized as follows. We first review the related work in Section 2. Section 3 presents the proposed double detectors-based GAN for malware generation. Section 4 describes the experiment and results. We conclude the paper with a conclusion and indicate possible research directions in Section 5.

\section{RELATED WORKS}
Hu et al \cite{7} introduced an algorithm based on GANs called MalGAN, which creates adversarial malware examples that can evade black-box machine-learning models. MalGAN uses a substitute detector to fit the black- box malware detection, and it is trained to minimize the detection probability of adversarial malware samples predicted by the substitute detector. MalGAN's advantage over traditional gradient-based adversarial sample generation algorithms is its ability to reduce the detection rate to nearly zero, making retraining-based defenses against adversarial samples less effective.

In the proposed Mal-LSGAN model to tackle these weaknesses. By using a least square (LS) loss function \cite{10} and new activation function combinations, Mal-LSGAN achieves a higher Attack Success Rate (ASR) and a lower True Positive Rate (TPR) in 6 ML detectors \cite{8}.

In the proposed AAGAN \cite{11}, an automated Android malware generation system based on Generative Adversarial Networks (GAN) that can successfully deceive current ML detectors. Their experiment results indicate that adversarial examples generated by their system can flip the prediction of the state-of-the-art detection algorithms in 99\% of cases using a real-world dataset. To defend against AE attacks, they improve the robustness of our detection system by alternatively retraining with these newly generated AEs. Surprisingly, after retraining five times, AAGAN can achieve an 89\% success rate in bypassing our malware detection system \cite{12}.

Shahpasand el al \cite{13} show a model for generating highly effective adversarial samples against machine learning-based malware detectors. The gap covered by this contribution is the threshold we provide on the maximum number of changes to an original malware sample. They used the Drebin dataset, choosing a subset of features that have the highest influence on classification with the same performance as the complete feature set. Their result shows that having a high performance for classifier accuracy on the original samples, the proposed model for generating adversarial samples can attack the classifiers with up to a 99\% success rate but only shows that the model can attack four different types of classifiers with a high chance of evasion.

\section{THE DOUBLE-DETECTORS BASED GAN FOR MALWARE GENERATION}
The proposed model includes 4 main blocks: generator, black-box detector, substitute detector and additional detector.
\begin{figure}[H]
    \centering
    \includegraphics[width=0.9\linewidth]{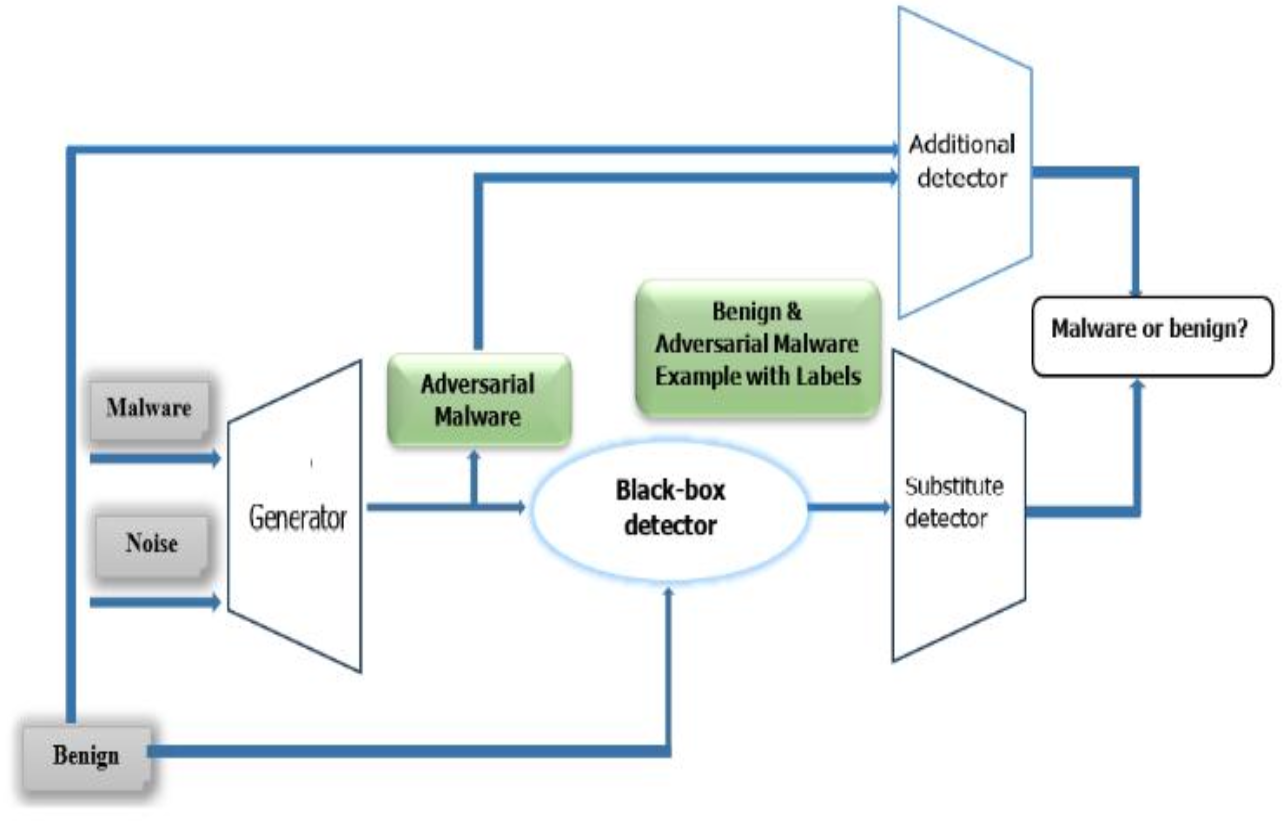}
    \caption{The architecture of Mal-D2GAN}
    \label{fig:The architecture of Mal-D2GAN}
\end{figure}

\subsection{Black-boxdetector}
This external system utilizes machine learning-based algorithms for malware detection. We assume that the malware author is only aware of the types of features used by the black-box detector. They do not know which machine learning algorithm is employed, nor do they have access to the trained model parameters. The malware author can receive the detection results of their programs from the black-box detector. In this model, the black-box detectors uses 8 classification algorithms such as Random forest (RF), Linear Regression (LR), Decision Tree (DT), Multilayer Perceptron (MLP), Support Vector Machine (SVM), AdaBoost (AB), GradientBoosting (GB), and K-nearest- Neighbor (KNN).
\subsection{Generator} 
This block generates adversarial malware data by transforming a malware feature vector into its adversarial sample. It requires a combination of the malware feature vector $\mathbf{m}$ and a noise vector $\mathbf{z}$ as input. Here, we use the design advantages of the DCGAN \cite{14} model for the generator block. The convolution layers in DCGAN are replaced with fully connected layers. Malware is an $M$-dimensional binary vector, with each element representing the presence or absence of a feature. $\mathbf{z}$ is a $Z$-dimensional vector with $Z$ being a hyperparameter. Each element of $\mathbf{z}$ is a random number sampled from a uniform distribution in the range $[0, 1)$. The purpose of $\mathbf{z}$ is to allow the generator to create diverse adversarial samples from a single malware feature vector. The structure of the generator is shown in Table~\ref{tab:generator_structure}.

\begin{table}[ht]
\centering
\caption{The generator (G) network structure}
\label{tab:generator_structure}
\begin{tabular}{|l|c|}
\hline
\textbf{Layer Type} & \textbf{Output Shape} \\
\hline
Input Layer (Malware) & (None, 160) \\\hline
Input Layer (Noise) & (None, 10) \\\hline
Concatenate (Malware + Noise) & (None, 170) \\\hline
LeakyReLU (Dense) & (None, 256) \\\hline
BatchNormalization & (None, 256) \\\hline
LeakyReLU (Dense) & (None, 256) \\\hline
BatchNormalization & (None, 256) \\\hline
LeakyReLU (Dense) & (None, 256) \\\hline
BatchNormalization & (None, 256) \\\hline
Sigmoid (Dense) & (None, 160) \\\hline
Maximum & (None, 160) \\
\hline
\end{tabular}
\end{table}

The input vector is fed into a multi-layer feed-forward neural network with weights $\theta_{g}$. The output layer of this network has $M$ neurons, and the activation function used by the last layer is sigmoid, which limits the output to the range $(0, 1)$. The hidden layers use LeakyReLU as the activation function due to its better performance and high learning cost parameters. The loss function used in the generator block is mean square error (MSE). The output of this generator network is denoted as $o$. 

Since the malware feature values are binary, a binary conversion process is applied to $o$ depending on whether the element values are greater than 0.5 or not, creating a binary vector ${o}{'}$. When creating adversarial samples for binary malware features, we exclusively focus on adding unrelated features to the original malware. The removal of any feature from the original malware could potentially disrupt its functionality. For example, if the API  {"DeleteFile"} is removed from a program, the program will not be able to perform normal deleting functions, and the malware may be detected. 

The non-zero elements of the binary vector ${o}{'}$ are used as unrelated features to add to the original malware. The final adversarial example can be represented as {$m' = m \mid${\textit{o}'}}, where ``$\mid$" is the element-wise binary OR operation. $m'$ is a binary vector, so the gradient cannot propagate back from the substitute detector to the generator. A smooth function $G$ is defined to receive gradient information from the substitute detector, as shown in:
\begin{equation}
\centering
\textit{G}_{\theta_g}(m, z)\ = max_(m, o)
\end{equation}

\subsection{Detectors}

\subsubsection{Substitute Detector}

Substitute detector is employed to approximate its behavior and supply gradient information for training the generator. Previously in , when there was no additional detector, substitute detector was a multi-layer feed-forward neural network with weights $\theta_d$ that received the program feature vector $\mathbf{m}$ as input. It classifies the program as either benign or malicious. We denote the probability of $\mathbf{x}$ being predicted as malware as $ D_{\theta_d}{(x)}$.

The substitute detector network plays a crucial role in enabling gradient-based adversarial training by serving as a differentiable approximation of the actual detection system.

\begin{table}[H]
\centering
\caption{The network structure of the substitute detector}
\label{tab:substitute_detector}
\begin{tabular}{|l|c|}
\hline
\textbf{Layer Type} & \textbf{Output Shape} \\
\hline
Input Layer (Malware) & (None, 160) \\
Sigmoid (Dense) & (None, 256) \\
Sigmoid (Dense) & (None, 1) \\
\hline
\end{tabular}
\end{table}

The training data for the substitute detector includes adversarial malware samples generated by the generator and benign samples from a benign dataset collected from various sources. Importantly, the training data for the substitute detector is all labeled by the black-box detector. The actual labels of the training data are not used to train the substitute detector. The goal of the substitute detector is to fit the black-box detector. 

The black-box detector will first detect this training data and label it as either malicious or benign. Then, these samples are used to train the substitute detector.

\subsubsection{Additional Detector}

In the new Mal-D2GAN model, an additional detector block is designed to work alongside the substitute detector to more effectively detect malware generated by the generator. Experimental results have demonstrated this effectiveness.

\begin{table}[H]
\centering
\caption{Network structure of the substitute detector}
\label{tab:substitute_detector_structure}
\begin{tabular}{|l|c|}
\hline
\textbf{Layer Type} & \textbf{Output Shape} \\
\hline
Input Layer (Malware) & (None, 160) \\\hline
Sigmoid (Dense) & (None, 256) \\\hline
Sigmoid (Dense) & (None, 1) \\
\hline
\end{tabular}
\end{table}

The data used to train additional detector will include
benign data from the training set as well as data generated
directly from the generator block. The input for both
substitute detector and additional detector is malware
represented as an M-dimensional vector. Substitute detector
and additional detector will be designed to work together to
predict whether the received data is malicious or benign.

\subsubsection{Adversarial example generation}
To train MalGAN, the malware author should first
collect two datasets: one containing malicious code samples
and another containing benign samples.

The combined loss function of substitute detector ($L_{D_1}$) and additional detector ($L_{D_2}$) is defined as in:

\begin{align}
L_{D_1} &= \frac{1}{2} \mathbb{E}_{x \in BB_{Benign}} \left[ \left( D_{1\theta_{d_1}}(x) \right)^2 \right] \nonumber \\
&\quad + \frac{1}{2} \mathbb{E}_{x \in BB_{Malware}} \left[ \left( D_{1\theta_{d_1}}(x) - 1 \right)^2 \right]
\end{align}
\begin{align}
L_{D_2} &= \frac{1}{2} \mathbb{E}_{x \in S_{Benign}} \left[ \left( D_{2\theta_{d_2}}(x) \right)^2 \right] \nonumber \\
&\quad + \frac{1}{2} \mathbb{E}_{x \in G_{Adversarial}} \left[ \left( D_{2\theta_{d_2}}(x) - 1 \right)^2 \right]
\end{align}
\begin{equation}
L_D = \alpha L_{D_1} + (1 - \alpha) L_{D_2}
\end{equation}

$BB_{\text{\textit{Benign}}}$ represents a set of programs identified as non-malicious by the black-box detector, and $BB_{\text{\textit{Malware}}}$ represents a set of programs detected as malware by the black-box detector. $S_{\text{\textit{Benign}}}$ is the original dataset of benign samples labeled to train the additional detector, and $G_{\text{\textit{Adversarial}}}$ refers to malware data generated by the generator for training the additional detector. To train the detectors, $L_D$ needs to be minimized with respect to the weights of the detectors. In (3), $\alpha$ is used to represent the importance of $L_{D_1}$ and $L_{D_2}$.

The loss function of the generator is defined as:

\begin{equation}
L_G = \frac{1}{2} \, \mathbb{E}_{m \in S_{\text{Malware}},\ z \sim \text{p}(0,1)} 
\left[ \left( D_{\theta_{d_1 d_2}} \left(G_{\theta_g}(m, z)\right) \right)^2 \right]
\end{equation}

\begin{algorithm}[H]
\caption{The Training Process of Mal-D2GAN}
\begin{algorithmic}[1]
\While{not converging}
    \State Sample a batch of malware ${M}$
    \State Generate adversarial samples ${M}'$ from Generator for ${M}$
    \State Sample a batch of benign samples ${B}$
    \State Label ${M}'$ and ${B}$ using black-box detector
    \State Label $S_{Benign}$ and $G_{Adversarial}$ using additional detector
    \State Update weights $\theta_{d1}$ and $\theta_{d2}$ of substitute detector and additional detector by descending along the gradients $\nabla_{\theta_1}{L}_{D1}$, $\nabla_{\theta_2} {L}_{D2}$
    \State Update weights $\theta_{g}$ of generator by descending along the gradients $\nabla_{\theta} {L}_G$
\EndWhile
\end{algorithmic}
\end{algorithm}

$S_{Malware}$ represents a dataset of actual malware and is not labeled by the black-box detector. ${L}_G$ is minimized with respect to the weights of the generator during training. Minimizing ${L}_G$ reduces the probability of the generated malware being predicted as harmful, thereby encouraging the substitute detector to classify malware as benign. Because the substitute detector aims to align with the black-box detector, training the generator will further deceive the black-box detector. The entire training process of Mal-D2GAN is illustrated in Algorithm 1.

\section{Experiments and Results}
\subsection{Experimental Setup}

The malware dataset used in the experiment was collected from various sources such as VirusShare \cite{15}, consisting of files in Portable Executable (PE) format. Each malware file, after being processed, is represented as a 160-dimensional API feature vector. 

The dataset contains 20{,}000 samples, of which 70\% are malware files and 30\% are benign files. In the experiment, the value of $\alpha$ was selected as 0.5.

To build the malware dataset, a behavior-based malware
analysis system called Cuckoo Sandbox \cite{16} is required. This system helps to analyze malware samples and provides
results in a JSON file format.

\begin{figure}[H]
    \centering
    \includegraphics[width=1\linewidth]{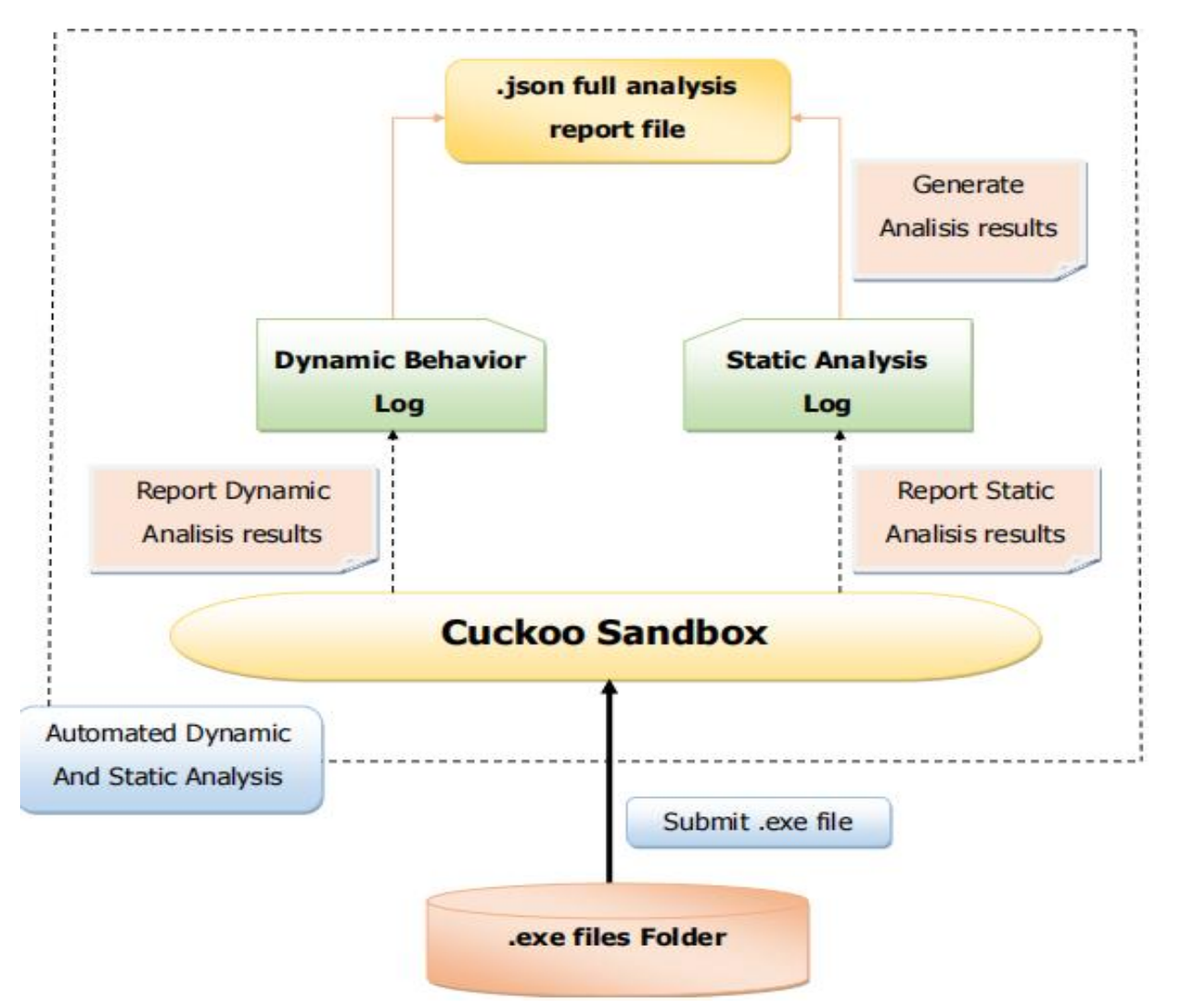}
    \caption{Collecting the report file from the Cuckoo sandbox}
    \label{fig:enter-label}
\end{figure}

To build the malware dataset, a behavior-based malware
analysis system called Cuckoo Sandbox \cite{16} is required. This system helps to analyze malware samples and provides
results in a JSON file format.

There are many features, including static features and
malicious features, but for the experiment, we will select
about 160 prioritized features. We use the Random Forest
classifier algorithm to determine the scores of the 160
dynamic features. 

The testing process of the Mal-D2GAN model uses a method for data splitting as follows: 80\% of the dataset is used as the training set, and 20\% of the dataset is used as the test set. The black-box detector block will use the training set data for training.

We conducted experiments on a computer with an Intel Core i5-12500H processor, 16GB of RAM, running Windows 11 and Anaconda \cite{17}, using 20 epochs fortraining and 5 epochs for retraining \cite{18}. 

\subsection{Experimental Results}

We will analyze the scenario where Mal-D2GAN and black-box detector use the same training set. In malware detection, the true positive rate (TPR) \cite{21} represents the rate of correctly detecting malware. After conducting adversarial attacks, the generated adversarial malware samples gain the ability to bypass detection algorithms. In the table below, a dataset with 20,000 samples is used to test this case \cite{19}.

\begin{table}[H]
\centering
\caption{True positive rate (in percentage) on original samples and adversarial examples Mal-D2GAN, Mal-LSGAN, MalGAN- on training set. “Adver.” represents adversarial examples}
\begin{tabular}{|c|c|c|c|c|c|c|}
\hline
\textbf{Classifier} & \multicolumn{2}{|c|}{\textbf{Mal-D2GAN}} & \multicolumn{2}{|c|}{\textbf{Mal-LSGAN}} & \multicolumn{2}{|c|}{\textbf{MalGAN}} \\\hline
\cline{2-7}
& Original & Adver & Original & Adver & Original & Adver \\
\hline
RF  & 97.25 & 2.28  & 97.75 & 20.14 & 97.02 & 31.68 \\\hline
LR  & 94.69 & 0.00  & 94.52 & 20.39 & 94.50 & 0.61  \\\hline
DT  & 97.35 & 0.00  & 94.64 & 0.98  & 97.40 & 6.60  \\\hline
SVM & 96.10 & 0.07  & 96.26 & 20.61 & 96.34 & 5.84  \\\hline
MLP & 98.31 & 0.16  & 97.80 & 20.93 & 97.11 & 35.09 \\\hline
AB  & 93.66 & 0.00  & 93.86 & 14.42 & 94.01 & 16.02 \\\hline
GB  & 97.35 & 0.37  & 97.35 & 14.38 & 97.40 & 49.86 \\\hline
KNN & 98.10 & 0.15  & 98.15 & 1.19  & 98.81 & 63.34 \\
\hline
\end{tabular}
\label{tab:results}
\end{table}

\begin{table}[H]
\centering
\caption{True positive rate (in percentage) on original samples and
adversarial examples when Mal-D2GAN, Mal-LSGAN and MalGAN on
the test set. “Adver.” represents adversarial examples}
\begin{tabular}{|c|c|c|c|c|c|c|}
\hline
\textbf{Classifier} & \multicolumn{2}{|c|}{\textbf{Mal-D2GAN}} & \multicolumn{2}{|c|}{\textbf{Mal-LSGAN}} & \multicolumn{2}{|c|}{\textbf{MalGAN}} \\\hline
\cline{2-7}
& Original & Adver & Original & Adver & Original & Adver \\
\hline
RF  & 94.74 & 2.38  & 98.16 & 20.59 & 97.10 & 31.80 \\\hline
LR  & 94.32 & 0.10  & 94.17 & 21.03 & 94.90 & 0.43  \\\hline
DT  & 94.45 & 0.10  & 93.80 & 1.20  & 97.36 & 6.33  \\\hline
SVM & 95.82 & 0.00  & 95.78 & 19.71 & 96.22 & 5.20  \\\hline
MLP & 96.48 & 0.18  & 96.22 & 20.33 & 97.54 & 37.15 \\\hline
AB  & 92.04 & 0.00  & 92.89 & 13.99 & 93.51 & 15.93 \\\hline
GB  & 96.11 & 0.18  & 96.00 & 14.07 & 97.06 & 51.33 \\\hline
KNN & 97.50 & 0.16  & 97.80 & 0.95  & 97.44 & 61.31 \\
\hline
\end{tabular}
\label{tab:test_results}
\end{table}

Overall, the results in Table~\ref{tab:results} and Table~\ref{tab:test_results} show that the Mal-D2GAN model reduced the detection accuracy (true positive rate) across 8 malware detection algorithms. The experimental results of the Mal-D2GAN model, along with comparisons with existing models such as Mal-LSGAN and MalGAN, demonstrate its effectiveness. It is evident that our Mal-D2GAN model surpasses the malware detection benchmarks with high performance metrics. Across various algorithms such as RF, LR, DT, SVM, MLP, AB, GB, and KNN, the true positive rate (TPR) mostly decreases to near 0\%. In comparison, for Mal-LSGAN, only DT approaches 0\%, while for MalGAN, only LR decreases to nearly 0\%. Specifically, the MLP classifier in the Mal-D2GAN model shows a dramatic drop in TPR from 98.31\% to 0.16\%. RF also decreases significantly to 2.28\%, and DT drops rapidly from 97.35\% to 0\%. 

These experimental results demonstrated the impact of the additional detector in the Mal-D2GAN model, which dramatically reduces the detection accuracy (true positive rate) in all 8 malware detectors.

\subsection{Retraining the Black-box Detector}

In this section, we will analyze the performance of Mal-D2GAN under the retraining-based defensive \cite{6} approach. If an anti-malware vendor collects enough adversarial malware examples, they can retrain the black-box detector on these adversarial examples to learn their signature and detect them.

We use RF, DT, AB, GB, and KNN as the black-box detectors due to their good performance. After retraining the black-box detector 5 times, it can detect all adversarial examples, as shown in the middle column, namely “Before retraining GAN model” \cite{20}.
\begin{table}[ht]
\centering
\caption{True positive rate (in percentage) on the adversarial examples after the black-box detector is retrained}
\begin{tabular}{|c|c|c|c|c|c|c|}
\hline
\textbf{Classifier} & \multicolumn{2}{|c|}{\textbf{Mal-D2GAN}} & \multicolumn{2}{|c|}{\textbf{Mal-LSGAN}} & \multicolumn{2}{|c|}{\textbf{MalGAN}} \\\hline
\cline{2-7}
& Before & After & Before & After & Before & After \\
\hline
RF  & 100 & 0.29  & 100 & 100   & 100 & 16.96 \\\hline
DT  & 100 & 0.88  & 100 & 20.07 & 100 & 36.90 \\\hline
AB  & 100 & 0.00  & 100 & 14.37 & 100 & 20.72 \\\hline
GB  & 100 & 0.00  & 100 & 100   & 100 & 43.61 \\\hline
KNN & 100 & 14.66 & 100 & 11.14 & 100 & 87.29 \\
\hline
\end{tabular}
\label{tab:retraining}
\end{table}

However, once antivirus vendors release the updated black-box detector publicly, malware authors will have access to it and can retrain Mal-D2GAN to target the new black-box detector. Following retraining, Mal-D2GAN generates new adversarial examples that remain undetected by RF, DT, AB, GB, and KNN, as indicated in the "After retraining model." 

Table~\ref{tab:retraining} shows the results of Mal-LSGAN after undergoing 5 rounds of retraining. It demonstrates that Mal-LSGAN performs well with DT, AB, and KNN. Meanwhile, the performance outcomes of MalGAN also plateaued at average effectiveness levels after the model retraining process.

\section{CONCLUSION AND FUTURE WORK}

In this paper, we develop a novel Mal-D2GAN model to enhance the robustness of the malware detector. The Mal-D2GAN architecture was designed with a double-detector and a least square loss function and tested on a dataset of 20,000 samples. The results show that the Mal-D2GAN model reduced the detection accuracy (true positive rate) in 8 malware detectors. This performance was then compared with that of the existing MalGAN and Mal-LSGAN models. The results also show that after retraining Mal-D2GAN, new adversarial examples remain undetected.

Additionally, detailed experiments demonstrate that combining the two detectors enhances the quality of adversarial sample generation. Moving forward, our goal is to further experiment with retraining using algorithms such as MLP, LR, and SVM, focusing on generating high-quality malware and testing it across various datasets.

\section*{ACKNOWLEDGMENT}

The authors would like to thank Khiem Phan Xuan for his contributions to carrying out the experiments for this project.

\bibliographystyle{IEEEtran} 
\bibliography{template.bib} 

\end{document}